\documentclass[10pt]{article}
\usepackage{epsf,amsmath,graphicx,citesort}

\newcommand{\figA}
{
\begin{figure}[t]
\begin{center}
\leavevmode
\vbox{
 \hbox{
 \epsfxsize=0.40\hsize \epsfbox{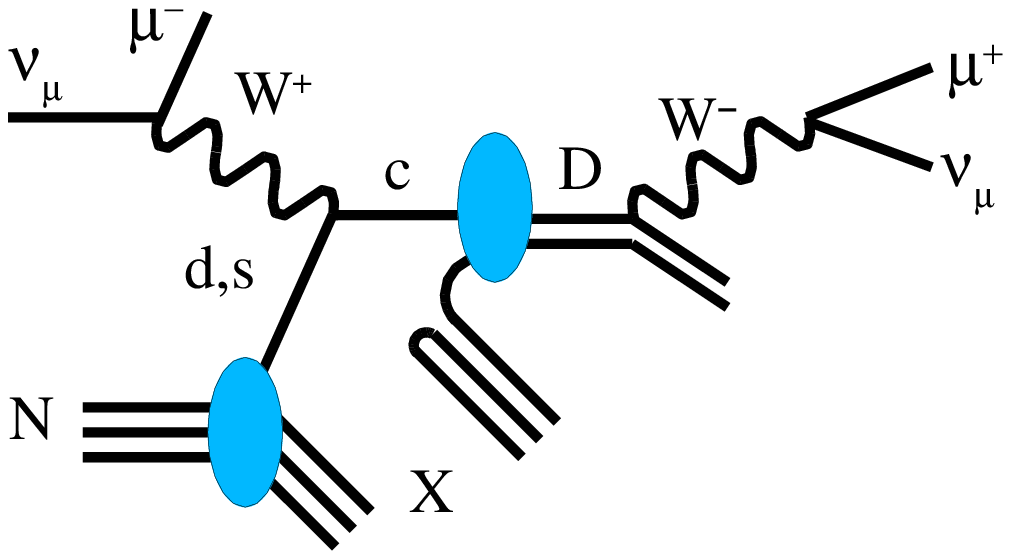}
 \epsfxsize=0.55\hsize \epsfbox{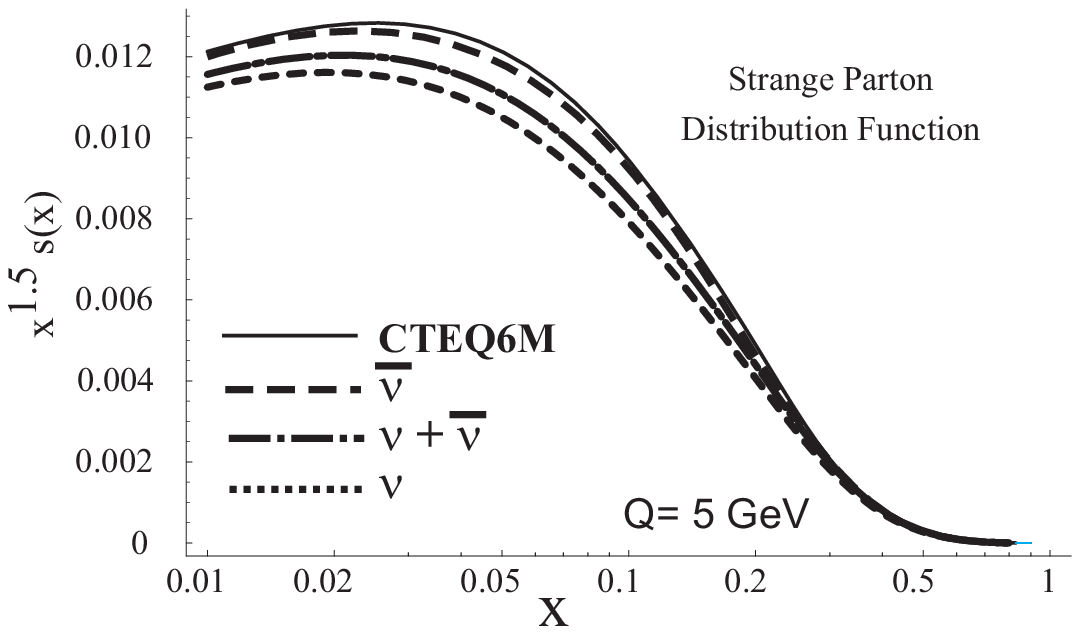} 
 }
}
\vskip -00pt
\end{center}
 \caption{
a) The basic process $\nu_\mu N \to \mu^- \mu^+ X$. 
\quad
b) The strange quark PDF vs. $x$
for the 4 PDF sets discussed in the text.
 \label{fig:A}
}
\end{figure}
}
\newcommand{\figB}
{
\begin{figure}[t]
\begin{center}
\leavevmode
\vbox{
 \hbox{
 \epsfxsize=0.45\hsize \epsfbox{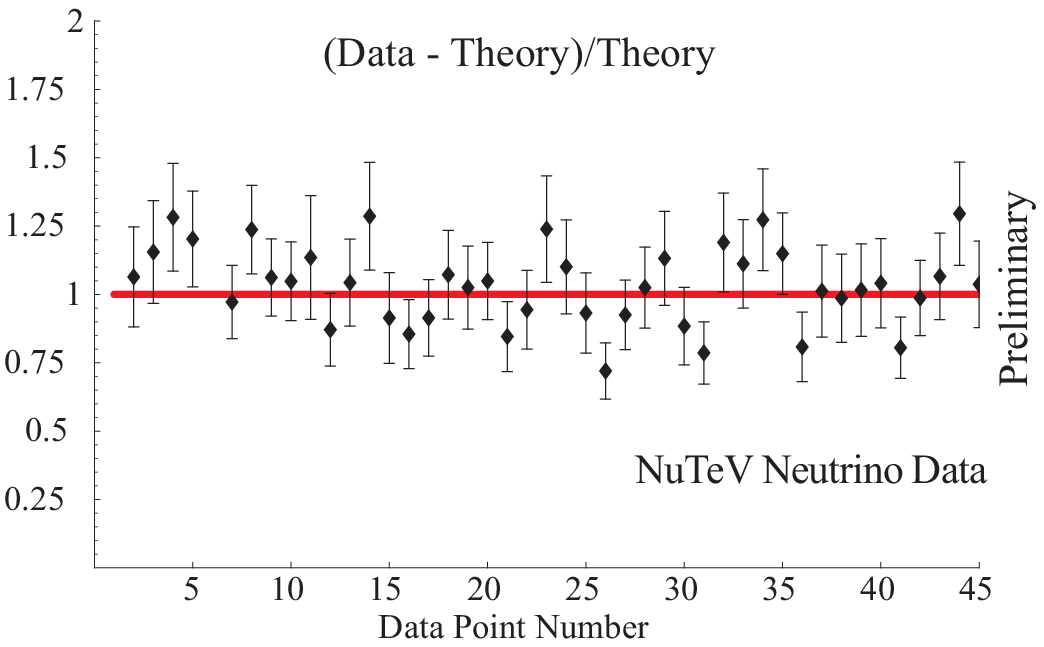} 
\hspace{15pt}
 \epsfxsize=0.45\hsize \epsfbox{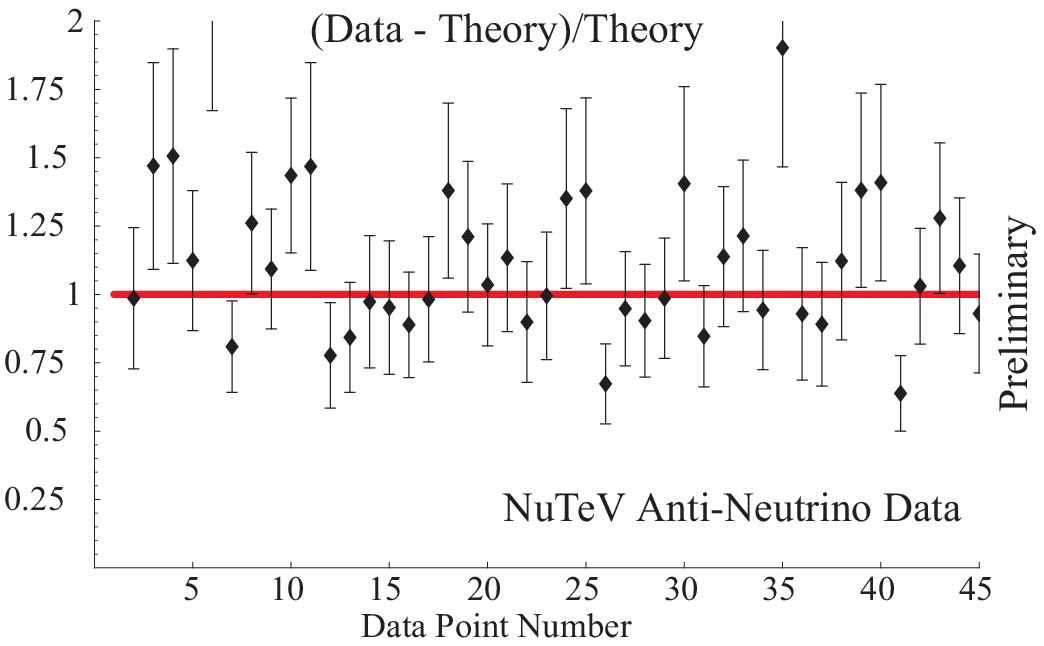} 
 }
}
\end{center}
 \vspace{-5ex}
 \caption{
Plot of $(Data - Theory)/Theory$ for the NuTeV $\nu$ and $\bar{\nu}$
data {\it vs.} data point number. 
Note the larger error bars on the $\bar{\nu}$ data due to the 
lower statistics. 
 \label{fig:B}
}
\end{figure}
}

\begin{document}

\title{
Neutrino Charm Production and Implications for PDF's
\thanks{
The work presented here is performed in collaboration with 
S.~Kretzer, 
P.~Nadolsky, 
W.K.~Tung
J.F.~Owens, 
S.~Kuhlmann,
J.~Pumplin, 
D.~Stump, 
M.H.~Reno,
J.~Huston, 
\&
H.~L.~Lai.}
\thanks{\it
To appear in the proceedings of 11th International 
Workshop on Deep Inelastic Scattering (DIS 2003), 
St. Petersburg, Russia, 23-27 Apr 2003.
}
}

\author{Fredrick I. Olness \\
Southern Methodist University, 
Dallas, TX USA 75275}

\maketitle

\begin{abstract}
\noindent
We have performed the first comprehensive global QCD analysis
including the CCFR and NuTeV di-muon data; this data provides strong
constraints on the strange quark PDF.

\end{abstract}

\subsection*{Introduction}

The recent measurements of both neutrino and anti-neutrino production
of charm (as di-muon final states) by the CCFR and NuTeV
collaborations provide important new information on the strange  quark
distribution, $s(x)$, of the nucleon.\cite{max,naples}
 We report here the first comprehensive global QCD analysis that
includes the CCFR and NuTeV di-muon data.

In previous global analyses, the predominant information on $s$ 
came from the difference of (large) inclusive cross sections
for neutral and charged current DIS; hence, the comparably small $s$
and $\bar{s}$ distributions extracted had large uncertainties.
Lacking better information, these studies often assumed the distribution was
of the form
$s(x)=\bar{s}(x) \sim \kappa (\bar{u}+\bar{d})/2$ with $\kappa \sim 0.5$. 
 The recent high-statistics measurements of 
$\sigma _{\nu N}^{\mu^{+}\mu ^{-}}$ and $\sigma _{\bar{\nu}N}^{\mu ^{+}\mu ^{-}}$ 
by the
CCFR and NuTeV experiments allow us to separately determine $s(x)$ and
$\bar{s}(x)$ with unprecedented accuracy.
 Neutrino induced di-muon production, $\nu/\bar{\nu} N \rightarrow
\mu^+ \mu^- X$, proceeds primarily through the subprocess $s \to c$ or
$\bar{s} \to \bar{c}$, and hence provides information on $s$ and $\bar{s}$
directly. (Cf., Fig~1a.)

We present a global analysis including this new di-muon data,
corrected for experimental cuts and efficiencies using information
provided by the experimental group.
 The new results are rich in physical content, in part, due to the
interplay of these high statistics measurements and the strong
constraints of the PQCD framework.

\figA

\subsection*{Global Analysis}

The inclusion of the CCFR-NuTeV neutrino and anti-neutrino di-muon production
data give direct handles on $s$ and $\bar{s}$ 
in the important $x$ range $\sim[0.01,0.2]$. 
 CCFR has recorded 5030 $\nu$ and 1060 $\bar{\nu}$ di-muon events, while
NuTeV has recorded 5012 $\nu$ and 1458 $\bar{\nu}$ di-muon events.
Additionally, NuTeV had a sign-selected beam to separate 
the $\nu$ and  $\bar{\nu}$ events. 
 The global QCD analysis contains the full data set from the CTEQ6
analysis, in addition to the di-muon data; as such, this can be
considered as an extension of the on-going CTEQ series of global
analysis.\cite{cteq6}

Incorporating this into the  global QCD analysis is a non-trivial task.
The experimental
measurement 
is the cross section for producing two final state muons
($d\sigma_{\mu^\pm \mu^\mp}$)  with experimental cuts, 
whereas the theoretical quantities that are
most directly related to the parton distribution analysis are the underlying
``charm quark production cross sections''  ($d\sigma_{c  \mu^\pm}$) 
for the process $\nu_\mu / \nu_{\bar{\mu}} N \to \mu^\mp c X$. 
We can relate these quantities via the equation: 
\begin{equation}
\left.
\frac{d\sigma_{\mu^\pm \mu^\mp} }{ dx \,dy } 
 = 
\int d\Gamma \, d\Omega \ 
\frac{d\sigma_{c  \mu^\pm} }{ dx \, dy \, d\Gamma} 
\, \otimes \, 
D_c(\Gamma)
\, \otimes \, 
\Delta_c(\Omega) \ 
\right|_{E_{\mu^\pm} > 5 \, GeV}
\end{equation}
Here,
$\Gamma$ and $\Omega$ denote the fragmentation 
and decay kinematic variables of the charmed quark 
and charmed hadron, respectively; 
$D_c(\Gamma)$ is the fragmentation function, and 
$\Delta_c(\Omega)$ is the decay distribution function.

The gap between these two cross sections are usually
bridged by Monte Carlo programs which incorporate experimental cuts and
efficiencies as well as fragmentation models. In our analysis, we
rely on a Pythia program provided by the CCFR-NuTeV 
collaboration.\footnote{%
We thank Tim Bolton and Max Goncharov, in particular, for providing this program,
as well as assistance to use it. Both are vital for carrying out this
project.} 
This Monte Carlo calculation in done in the spirit and the framework
of leading-order (LO) QCD. Accordingly, the theoretical formulas used are also
in LO. Since charm production has such a small cross section in the
experimental kinematic range covered, this approximation is perfectly adequate
(as we will verify later) 
for a first study of the di-muon data within the global QCD analysis framework.
Needless to say, all fully inclusive (large) cross sections used in this
study are treated in NLO QCD, as in all modern global analyses.

\figB

\subsection*{Results}

We parameterize the strange quark in the form:
\begin{equation}
s(x,Q_0) \equiv \bar{s}(x,Q_0)
= a_0 \, x^{a_1} (1-x)^{a_2} e^{a_3 x} 
\left( 
1+e^{a_4}x
\right)^{a_5}
\end{equation}
In the present analysis, we take $s(x) \equiv \bar{s}(x)$; 
in later studies we shall consider a non-symmetric 
strange distribution.\cite{nutev,bpz,p1,p2}

We will consider four separate sets of PDF's:
the CTEQ6M set (which is {\it not} fit to the di-muon data), 
a {\it Constrained} fit where only the normalization 
$\{a_0 \}$  is allowed to vary; 
a {\it Mixed} fit where both  $\{a_0,a_2 \}$  are allowed to vary; 
a {\it Free} fit  where all   $\{a_i \}$  are allowed to vary;

The CTEQ6M PDF's yield a respectable 
$\chi^2$ of 2173 for 1991 data points (CTEQ6M was not fit with the di-muons), 
and
the {\it Constrained}, {\it Mixed}, and {\it Free} fits  
yield 2144, 2142, and 2133, respectively. 
While it is reassuring to see the $\chi^2$ decrease as we increase the 
number of free parameters, 
it will require additional study to determine whether
the data can actually constrain all these parameters.

It is also interesting to note that as we move from the CTEQ6M fit
to the fits including the di-muons, the $\chi^2/DOF$ for all the experiments
(excluding the di-muons) are generally unchanged to within a percent.
From this observation, it is evident that the effect of adding in the 
di-muon data serves to adjust the strange PDF, but has virtually no effect
on the other data sets. Or, equivalently, this demonstrates that 
the di-muon data plays a dominant role in determining the strange distribution.

The observation that it is primarily the di-muon data which influence the 
strange distribution prompts us to fit the  neutrino and anti-neutrino 
data separately to investigate the extent to which this data can 
determine the $s(x)$ and  $\bar{s}(x)$ distributions independently.\footnote{
A complete analysis which fits  the  neutrino and anti-neutrino 
data simultaneously and 
allows $s(x \not= \bar{s}(x)$ is in progress.\cite{p1,p2}}
At leading-order, 
the neutrino induced process 
$\nu s \to \mu^- c$ is tied to the strange distribution, while  
the anti-neutrino induced process 
$\bar{\nu} \bar{s} \to \mu^+ \bar{c}$ is tied to the anti-strange 
distribution. 
Of course, these relations will be complicated at NLO; however it
is useful to see if, and how, these separate data sets pull the fit. 

In Fig.1b we display the obtained strange distribution for the
combined $\nu$ and $\bar{\nu}$ fit, as well as the individual fits. 
We observe that the  $\nu$  data yields a smaller PDF, while the 
$\bar{\nu}$ yields a slightly larger fit than the combined set. 
Note that this effect is consistent with the results of the NuTeV
fit by Goncharov,\cite{max} which obtained 
$\kappa = 0.35$ for  the $\nu$ data and 
$\bar{\kappa} = 0.41$ for  the $\bar{\nu}$ data. 

This result suggests that the di-muon data is capable of  
providing information about the $s$ and $\bar{s}$ distributions separately. 
However, to properly extract this information requires a simultaneous 
fit to the  $\nu$ and $\bar{\nu}$ while imposing the strangeness sum 
rule on the PDFs: $\int dx \, s(x) - \bar{s}(x) = 0$.

\subsection*{Conclusion}

We present the first global analysis which includes the CCFR and NuTeV
neutrino and anti-neutrino di-muon production data.
We find several classes of solutions in the strangeness sector that
are consistent with all relevant world data used in these global
analyses. 

While the CTEQ6M PDF provided a good description of the data, we
obtained an improved fit if we free the parameters of the strange
quark PDF to allow them to conform to the di-muon data.
 A more comprehensive study of this data is in progress,\cite{p1,p2} and 
future work includes relaxing the constraint of $s =
\bar{s}$, and including the NLO modeling of the charm
production cross section.


We thank members of CCFR and NuTeV collaboration for discussions about
their experimental results.  In particular, we thank T. Bolton for
assistance with the di-muon Fortran programs.  F.O. acknowledges the
hospitality of MSU and BNL where a portion of this work was performed.
This research was supported by the U.S. Department of Energy (Contract
No.~DE-FG03-95ER40908), and by the Lightner-Sams Foundation.

\def\bib#1{\bibitem{#1}}  


\end{document}